\documentclass[aps,prd,groupedaddress,amssymb,twocolumn,eqsecnum,showpacs,epsfig,nofootinbib]{revtex4}

\usepackage{graphicx}          
\usepackage{bm}
\usepackage{dcolumn}
\usepackage{amsmath}       
  
\usepackage{amsmath}
\usepackage{amssymb} 

\numberwithin{equation}{section}
\usepackage{epsfig}

\begin{document}
\newcommand{\newc}{\newcommand}

\newc{\be}{\begin{equation}}
\newc{\ee}{\end{equation}}
\newc{\bear}{\begin{eqnarray}}
\newc{\eear}{\end{eqnarray}}
\newc{\bea}{\begin{eqnarray*}}
\newc{\eea}{\end{eqnarray*}}
\newc{\D}{\partial}
\newc{\ie}{{\it i.e.} }
\newc{\eg}{{\it e.g.} }
\newc{\etc}{{\it etc.} }
{\newc{\etal}{{\it et al.}}
\newc{\lcdm}{$\Lambda$CDM}
\newcommand{\nn}{\nonumber}
\newc{\ra}{\rightarrow}
\newc{\lra}{\leftrightarrow}
\newc{\lsim}{\buildrel{<}\over{\sim}}
\newc{\gsim}{\buildrel{>}\over{\sim}}
\newcommand{\mincir}{\raise
-3.truept\hbox{\rlap{\hbox{$\sim$}}\raise4.truept\hbox{$<$}\ }}
\newcommand{\magcir}{\raise
-3.truept\hbox{\rlap{\hbox{$\sim$}}\raise4.truept\hbox{$>$}\ }}

\title{Cosmological equivalence between the Finsler-Randers space-time 
and the DGP gravity model}

\author{Spyros Basilakos}\email{svasil@academyofathens.gr}
\affiliation{Academy of Athens, Research Center for Astronomy and
Applied Mathematics,
 Soranou Efesiou 4, 11527, Athens, Greece}

\author{Panayiotis Stavrinos}\email{pstavrin@math.uoa.gr}
\affiliation{Department of Mathematics
University of Athens, Panepistemiopolis, Athens 157 83, Greece}
  
\begin{abstract}
We perform a detailed comparison between the Finsler-Randers cosmological 
model and the Dvali, Gabadadze and Porrati braneworld model.
If we assume that the spatial curvature is strictly equal to zero then 
we prove the following interesting proposition: 
{\it despite the fact that the current 
cosmological models have a completely different geometrical origin 
they share exactly the same Hubble expansion}. This 
implies that the Finsler-Randers model is cosmologically equivalent with that 
of the DGP model as far as the cosmic expansion is concerned.
At the perturbative level we find that the Finsler-Randers growth index of 
matter perturbations is $\gamma_{FR} \simeq 9/14$ which is 
somewhat lower than that of DGP 
gravity ($\gamma_{DGP} \simeq 11/16$) implying
that the growth factor of the Finsler-Randers model is 
slightly different ($\sim 0.1-2\%$) from the one 
provided by the DGP gravity model.

\end{abstract}
\maketitle

\section{Introduction}
Geometrical dark energy models act as an important alternative to 
the scalar-field dark energy models, since they can explain 
the accelerated expansion of the universe. 
Such an approach is an attempt to evade the
coincidence and cosmological constant problems of 
the standard $\Lambda$CDM model.
In this framework, one may consider that the dynamical effects
attributed to dark energy can be resembled by the effects of a nonstandard
gravity theory implying that the present accelerating stage of the universe
can be driven only by cold dark matter, under a modification of the nature of
gravity. 

Particular attention over the last decade has been paid on the 
so-called Finsler-Randers (hereafter FR) cosmological model \cite{Rad41}.
In general metrical extensions of Riemann geometry can provide a Finslerian 
geometrical structure in a manifold which leads to generalized 
gravitational field theories. During the last decade there is a rapid 
development of applications of Finsler geometry in its FR context, mainly in
the topics of general relativity, astrophysics and  
cosmology \cite{Rad41,Goe99,Stav04,Per06,Bog08,Stav07,Gib07,Chang,Kou09,Kou10,Ska10,Ska11,Kost11,Mav11,Mav12,Vac12a,Vac12b,Kost12,Wer12,Stav12}.
It has been 
found \cite{Stav07} that the FR field equations 
provide a Hubble parameter that contains 
an extra geometrical term which   
can be used as a possible candidate for dark energy.

Of course, there are many other possibilities to explain the present
accelerating stage. Indeed, in the literature one can find a 
large family of modified gravity models (for review see Refs. \cite{Ame10,Ame10b})
which include the braneworld 
Dvali, Gabadadze and Porrati (hereafter DGP; \cite{DGP}) model, 
$f(R)$ gravity theories \cite{Sot10}, scalar-tensor theories \cite{scal}
and Gauss-Bonnet gravity \cite{gauss}. 
Technically speaking, it would be interesting if we could find a way to unify 
(up to a certain point) the geometrical dark energy models at the 
cosmological level. In general, we would like to pose the following
question: {\it how many (if any) of the above geometrical dark energy models
can provide exactly the same Hubble expansion?} 
In the current work we prove that the flat FR and DGP models respectively
share the same Hubble parameter which means that  
the two geometrical models are cosmologically equivalent, as far as the cosmic 
expansion is concerned.

The structure of the paper is as follows. 
Initially in Sec. II, we briefly discuss the DGP gravity model, 
while in Sec. III we present the main properties of the FR model.
In Sec. IV, we 
study the linear growth of perturbations and
constrain the FR growth index. 
Finally, in Sec. VI we summarize the basic results.

\section{The DGP cosmological model}
In this section, we briefly describe the main features of the DGP gravity 
model. The idea here is that the
''accelerated'' expansion of the universe can be explained by a
modification of the gravitational interaction in which gravity
itself becomes weak at very large distances (close to the Hubble
scale) due to the fact that our four-dimensional lives on 
a five-dimensional manifold \cite{Deff,Wei08}. Note that the Einstein field
equations are defined on the five-dimensional brane. In this 
framework, the modified Friedmann equation can be written as:
\be
\label{DGP1}
H^{2}+\frac{k}{a^{2}}-\frac{2M^{3}_{(5)}}{M^{2}_{(4)}}\left( H^{2}+\frac{k}{a^{2}}\right)^{1/2}=
\frac{8\pi G}{3}\rho_{m} 
\ee
where $a(t)$ is the scale factor of the universe and
$H(t)\equiv\dot{a}/a$ is the Hubble function
and $k=0,\pm1$ is the spatial
curvature parameter. 

Notice that $M_{(5)}$ and $M_{(4)}$ are the 5D and 4D Planck 
masses respectively. 
Inserting the following present-value quantities into Eq.(\ref{DGP1}) 
\be
\label{OO1}
\Omega_{rc}=\frac{1}{4r^{2}_{c}H^{2}_{0}}, \;\;\Omega_{k0}=-\frac{k}{H^{2}_{0}},\;\;
\Omega_{m0}=\frac{8\pi G \rho_{m0}}{3H^{2}_{0}}
\ee
one can write 
\be 
E^{2}(a)=\left[\sqrt{\Omega_{m0} a^{-3}+\Omega_{rc}}+\sqrt{\Omega_{rc}}\right]^{2}
+\Omega_{k0}a^{-2}
\ee
where $r_{c}=M^{2}_{(4)}/2M^{3}_{(5)}$ and 
$E(a)=H(a)/H_{0}$. Using a spatially flat geometry ($\Omega_{k0}=0$) 
and $E(1)=1$ the above normalized 
Hubble parameter takes the form 
\be
\label{nfe2}
E^{2}(a)= \Omega_{m0}a^{-3}+\Delta H^{2}.
\ee
where the quantity $\Delta H^{2}$ is given by
\be
\label{anfe2}
\Delta H^{2}=2\Omega_{rc}+2\sqrt{\Omega_{rc}}
\sqrt{\Omega_{m0}a^{-3}+\Omega_{rc}}
\ee
with $\Omega_{rc}=(1-\Omega_{m0})^{2}/4$. 
On the other hand Linder and Jenkins \cite{Linjen03} have shown 
that the corresponding effective (geometrical) 
dark energy equation of state parameter of Eq.(\ref{nfe2}) is written as 
\begin{equation}
\label{eos222}
w(a)=-1-\frac{1}{3}\;\frac{d{\rm ln}\Delta
H^{2}}{d{\rm ln}a}.
\end{equation}
Therefore, from Eq.(\ref{eos222}), it is
easily shown that the geometrical 
dark energy equation of state parameter of the flat DGP model reduces to 
\be
w(a)=-\frac{1}{1+\Omega_{m}(a)} 
\ee
where
\be 
\label{ddomm}
\Omega_{m}(a)=\frac{\Omega_{m0}a^{-3}}{E^{2}(a)} \;.
\ee

From the observational point of view the DGP gravity model has been 
tested well against the available cosmological data 
\cite{Fair06,Maa06,Alam06,Song07}. Although 
the flat DGP model was found to be consistent with SNIa data, 
it is under the observational pressure by including
in the statistical analysis the data of the 
baryon acoustic oscillation (BAO) and the 
cosmic microwave background (CMB) shift parameter \cite{Maa06}. Furthermore, 
it has been found (cf. Ref. \cite{Song07}) that the 
integrated Sachs-Wolfe (ISW) effect 
poses a significant problem for the DGP cosmology, 
especially at the lowest multipoles.

\section{The Finsler-Randers type cosmology}
The FR cosmic scenario is based on the Finslerian geometry 
which extends the traditional Riemannian geometry.
Notice that a Riemannian geometry is also a Finslerian.
Bellow we discuss only the 
main features of the theory (for more details see \cite{Rund,Mir,Bao,VV}).
Generally, a Finsler space is derived from a generating differentiable 
function $F(x,y)$ on a tangent bundle $F:$$TM \rightarrow R$, 
$TM=\tilde{T}(M)/\{0\}$ on a manifold $M$. The function $F$ is a degree-one 
homogeneous function with respect to $y=\frac{dx}{dt}$ and it is continuous 
in the zero cross section. In other words, $F$ introduces a structure 
on the space-time manifold $M$ that is called Finsler space-time.
In the case of a FR space-time we have 
\be 
\label{FF}%
F(x,y)=\sigma(x,y)+u_{\mu}(x)y^{\mu}, \;\; \sigma(x,y)\equiv \sqrt{a_{\mu \nu}y^{\mu}y^{\nu}}
\ee
where $a_{\mu \nu}$ is a Riemannian metric and $u_{\mu}=(u_{0},0,0,0)$ is 
a weak primordial vector field with $|u_{\mu}| \ll 1$. 
Now the Finslerian metric tensor $f_{\mu \nu}$ is constructed by the 
Hessian of $F$
\begin{eqnarray}
f_{\mu \nu}=\frac{1}{2} \frac{\partial^{2} F^{2}}{\partial y^{\mu} \partial y^{\nu}} \label{Hes} \;.
\end{eqnarray}
It is interesting to mention that the Cartan tensor 
$C_{\mu \nu k }=\frac{1}{2} \frac{\partial f_{\mu \nu}}{\partial y^{k}}$ 
is a significant ingredient of the Finsler geometry. Indeed it has been 
found \cite{Stav07} that $u_{0}=2C_{000}$. 

Armed with the above, the FR field equations are given by
\begin{equation}
L_{\mu\nu}=8\pi G \left(T_{\mu \nu}-\frac{1}{2}Tg_{\mu\nu}\right)\;\;\;\;c\equiv 1
\label{EE}%
\end{equation}
where $L_{\mu\nu}$ is the Finslerian 
Ricci tensor, $g_{\mu \nu}=Fa_{\mu \nu}/\sigma$, $T_{\mu\nu}$ is the 
energy-momentum tensor and $T$ is the trace of the 
energy-momentum tensor. 
Modeling the expanding universe as a Finslerian 
perfect fluid that includes radiation and matter 
with four-velocity $U_{\mu}$ for comoving observers\footnote{Here we use
$U^{\alpha}=\frac{dx^{\alpha}}{dt}=y^{\alpha}=(1,0,0,0)$, where $t$ is 
the cosmic time.}, we have $T_{\mu\nu}=-P\,f_{\mu\nu}+
(\rho+P)U_{\mu}U_{\nu}$, where $\rho=\rho_{m}+\rho_{r}$ and $P=P_{m}+P_{r}$ 
are the total energy density and pressure of the cosmic fluid
respectively. 
Note that $\rho_{m}=\rho_{m0}a^{-3}$ is the matter density, 
$\rho_{r}=\rho_{r0}a^{-4}$ denotes the density of the radiation and
$P_{m}\equiv 0$, $P_{r}\equiv \rho_{r}/3$ 
are the corresponding pressures\footnote{We use the fact that the radiation 
component is negligible in the matter-dominated era.}. 
Thus the energy-momentum tensor becomes 
$T_{\mu\nu}={\rm diag}\left(\rho,-Pf_{ij}\right)$, where the Greek indices belong 
to ${0,1,2,3}$ and the Latin ones to ${1,2,3}$.

In the context of a FLRW metric\footnote{The nonzero components of the 
Finslerian Ricci tensor are: $L_{00}=3(\frac{\ddot{a}}{a}+3\frac{\dot{a}}{4a}\dot{u}_{0})$ and 
$L_{ii}=-(a\ddot{a}+2\dot{a}^{2}+2k+\frac{11}{4}a\dot{a}\dot{u}_{0})/\Delta_{ii}$ where
$(\Delta_{11},\Delta_{22},\Delta_{33})=(1-kr^{2},r^{2},r^{2}{\rm sin}^{2}\theta)$.} 
\begin{eqnarray}
a_{\mu \nu}={\rm diag}\left(1,-\frac{a^2}{1-kr^2},-a^2r^2,-a^2r^2{\rm sin}^2\theta \right)
\label{SF.1}%
\end{eqnarray}
the gravitational FR field equations (\ref{EE}), for comoving observers, 
boil down to modified Friedmann's equations \cite{Stav07}
\be
\label{FRF11}%
\frac{\ddot{a}}{a}+\frac{3}{4}\frac{\dot{a}}{a}Z_{t}=-\frac{4\pi G}{3}(\rho+3P)
\ee
\be
\label{FRF22}%
\frac{\ddot{a}}{a}+2\frac{\dot{a}}{a}+2\frac{k}{a^{2}}+
\frac{11}{4}\frac{\dot{a}}{a}Z_{t}=4\pi G(\rho-P)
\ee
where the over-dot denotes derivative with respect to the cosmic time $t$
and $Z_{t}=\dot{u}_{0}<0$ (see Ref. \cite{Stav07}). 
With the aid of the Eqs. (\ref{FRF11}) and (\ref{FRF22}) we 
obtain, after some simple algebra, the Friedman-like expression in the matter
dominated era ($\rho=\rho_{m}$)
\begin{eqnarray}
\label{6} 
H^2+\frac{k}{a^2}+HZ_{t}=\frac{8\pi G}{3}\rho_{m} 
\end{eqnarray}
which looks similar to the form of
the DGP Friedmann equation [see Eq.(\ref{DGP1}].
Obviously, the extra term $H(t)Z_{t}$ in the modified Friedmann 
equation (\ref{6}) affects the dynamics of the universe.
If we consider $u_{0}\equiv 0$ (or $C_{000}\equiv 0$, $F/\sigma=1$), 
which implies $Z_{t}=0$, 
then the field equations (\ref{EE}) reduce
to the nominal Einstein's equations ($L_{\mu \nu}=R_{\mu \nu}$, where 
$R_{\mu \nu}$ is the usual Ricci tensor) a solution of which is the 
usual Friedman equation.

Therefore, utilizing the last two equalities
of Eq.(\ref{OO1}), Eq.(\ref{6}) and $E(a)=H(a)/H_{0}$ one can 
easily show that the normalized Hubble parameter is written as: 
\begin{eqnarray}
\label{EE11}
E^{2}(a)=\left[\sqrt{\Omega_{Z_{t}}+\Omega_{m0}a^{-3}+\Omega_{k0}a^{-2}}+\sqrt{\Omega_{Z_{t}}} \right]^{2}
\end{eqnarray}
where $\sqrt{\Omega_{Z_{t}}}=-\frac{Z_{t}}{2H_{0}}$. Assuming now
a spatially flat geometry $k=0$ ($\Omega_{k0}=0$) and $E(1)=1$, 
the above expression becomes
\be
\label{nfe22}
E^{2}(a)= \Omega_{m0}a^{-3}+\Delta H^{2}_{FR}.
\ee
where $\Delta H^{2}_{FR}$ is given by:
\be
\Delta H^{2}_{FR}=2\Omega_{Z_{t}}+2\sqrt{\Omega_{Z_{t}}}
\sqrt{\Omega_{m0}a^{-3}+\Omega_{Z_{t}}}
\ee
with $\Omega_{Z_{t}}=(1-\Omega_{m0})^{2}/4$. Amazingly, 
the Hubble parameter of the FR cosmology 
reduces to that of the flat DGP gravity, $\Delta H^{2}_{FR}=\Delta H^{2}$ 
[see Eqs. (\ref{nfe2}) and \ref{anfe2} or Eqs.(\ref{DGP1}) and (\ref{6})].

The importance of the current work is that we find that 
the flat FR model has exactly the same Hubble parameter as
the flat DGP gravity model, despite the fact that the geometrical base of
the two models is completely different. Our result implies that 
the flat FR and the DGP models 
can be seen as equivalent cosmologies 
as far as the Hubble expansion is concerned. Below we 
investigate at the perturbative 
level the predictions 
of the FR model with the DGP cosmology in order to show
the extend to which they are comparable.
 
\section{The linear matter fluctuations}
In this section, we briefly present 
the basic equation which governs the behavior of the matter
perturbations on subhorizon scales and within the context of any 
dark energy 
model, including those of modified gravity (``geometrical dark energy''), 
in which the dark energy is homogeneously distributed. 
The reason for investigating the 
growth analysis in this work is to give the reader the 
opportunity to appreciate also 
the relative strength and similarities
of the FR and DGP models at the perturbative level.

At subhorizon scales the effective (geometrical in our case) 
dark energy component 
is expected to be smooth and thus it is
fair to consider perturbations only on the matter component of the
cosmic fluid \cite{Dave02}. The evolution equation
of the matter fluctuations $\delta_{m}\equiv \delta\rho_m/\rho_m$, for 
cosmological models where the dark energy
fluid has a vanishing anisotropic stress and the matter fluid is not
coupled to other matter species 
(see Refs. \cite{Lue04},\cite{Linder05},\cite{Stab06},\cite{Uzan07},\cite{Linder2007},\cite{Tsu08},\cite{Gann09}), is given by:
\be
\label{odedelta} 
\ddot{\delta}_{m}+ 2H\dot{\delta}_{m}=4 \pi G_{\rm eff} \rho_{m} \delta_{m} \;.
\ee
where $G_{\rm eff}$ is the effective Newton's constant and 
$\rho_{m}$ is the matter density. 
Transforming Eq.(\ref{odedelta})
from $t$ to $a$ ($\frac{d}{dt}=H\frac{d}{d\ln a}$),
we simply obtain 
\be
\label{dela}
\frac{a^{2}}{\delta_{m}}\frac{d^{2}\delta_{m}}{da^{2}}+
\left(3+\frac{d{\rm ln}E}{d{\rm ln}a}\right)\frac{a}{\delta_{m}}
\frac{d\delta_{m}}{da}=
\frac{3}{2}\Omega_{m}(a)\frac{G_{\rm eff}(a)}{G_{N}} \;.
\ee
with $G_{N}$ denoting Newton's gravitational constant.
It is interesting to mention that
solving Eq.(\ref{dela}) for
the concordance $\Lambda$ cosmology \footnote{For the usual $\Lambda$CDM
cosmological model we have
$w(a)=-1$, $\Omega_{\Lambda}(a)=1-\Omega_{m}(a)$ and $G_{\rm eff}(a)=G_{N}$.}, we 
derive the well-known perturbation growth factor \cite{Peeb93}
scaled to unity at the present time
\be\label{eq24}
\delta_{m} \propto D(z)=\frac{5\Omega_{m0}
  E(z)}{2}\int^{+\infty}_{z}
\frac{(1+u)du}{E^{3}(u)} \;\;.
\ee
Notice that we have used $a(z)=1/(1+z)$.

At this point we define the so-called growth rate of clustering
which is an important parametrization
of the matter perturbations \cite{Peeb93}
\be
\label{fzz221}
f(a)=\frac{d\ln \delta_{m}}{d\ln a}\simeq \Omega^{\gamma}_{m}(a) \;.
\ee
The parameter $\gamma$ is the growth index
which plays a significant role in cosmological studies 
(see Refs.~\cite{Silv94,Wang98,Linjen03,Lue04,Linder2007,Nes08}).

Combining the first equality of 
Eq.(\ref{fzz221}) with Eq.~(\ref{dela}), we derive (after 
some algebra) that  
\be
\label{fzz222}
\frac{df}{d\Omega_{m}}\frac{d\Omega_{m}}{d{\rm ln}a}+f^{2}+\left(2+\frac{d{\rm ln}E}{d{\rm ln}a}\right)f
= \frac{3}{2}\Omega_{m}(a)\frac{G_{\rm eff}(a)}{G_{N}} \;.
\ee

In our case the basic quantities of Eq.(\ref{fzz222}) are 
(see also \cite{Lue04,Gong10})
\begin{equation}
\label{F11}
\frac{d{\rm ln}E}{d{\rm ln}a}=\left\{ \begin{array}{cc}
       -\frac{3\Omega_{m}(a)}{1+\Omega_{m}(a)}  &\;\;
       \mbox{DGP or FR}\\
       -\frac{3}{2}\Omega_{m}(a) & \mbox{$\Lambda$CDM}
       \end{array}
        \right.
\end{equation}

\begin{equation}
\label{G11}
\frac{d{\rm ln}\Omega_{m}}{d{\rm ln}a}=\left\{ \begin{array}{cc}
       -\frac{3\Omega_{m}(a)\left[1-\Omega_{m}(a)\right]}{1+\Omega_{m}(a)}  &\;\;
       \mbox{DGP or FR}\\
-3\Omega_{m}(a)\left[1-\Omega_{m}(a)\right]  & \mbox{$\Lambda$CDM}
       \end{array}
        \right.
\end{equation}
and 

\begin{equation}
\label{OF11}
\frac{G_{\rm eff}(a)}{G_{N}}=\left\{ \begin{array}{cc}
       1 &
       \mbox{$\Lambda$CDM or FR}\\
\frac{2+4\Omega^{2}_{m}(a)}{3+3\Omega^{2}_{m}(a)} & \mbox{DGP.}
       \end{array}
        \right.
\end{equation}

Inserting the ansatz 
$f\simeq \Omega^{\gamma(\Omega_{m})}_{m}$ into 
Eq. (\ref{fzz222}), using simultaneously Eqs. (\ref{F11}), (\ref{G11}), 
(\ref{OF11}) and 
performing a first order Taylor expansion
around $\Omega_{m}=1$ (for a similar analysis see Refs. \cite{Linder2007, Nes08, Bas12}) 
we find that the asymptotic value of the FR growth index to 
the lowest order is $\gamma_{FR}\simeq 9/14$, while
in the case of the DGP 
braneworld model we have $\gamma_{DGP} \simeq 11/16$ 
(see also \cite{Linder2007,Gong10,Wei08,Fu09}). Notice that for
the concordance $\Lambda$CDM cosmology it has been found 
(see \cite{Silv94},\cite{Wang98},\cite{Linder2007},\cite{Nes08})
that  $\gamma_{\Lambda} \simeq 6/11$. Using the above we find the 
following restriction 
$$
\gamma_{\Lambda} <\gamma_{FR} <\gamma_{DGP} \;\;.
$$  
The small difference ($\sim 7\%$) between $\gamma_{FR}$ 
and $\gamma_{DGP}$ is due to $G_{\rm eff}/G_{N}$ used 
in the growth analysis (see Eq.\ref{OF11}). 

\begin{figure}[ht]
\mbox{\epsfxsize=9cm \epsffile{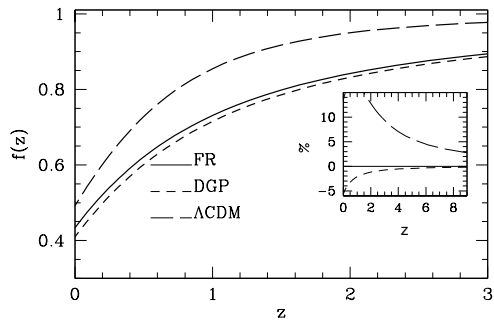}}
\caption{The evolution of the growth rate of clustering $f(z)$.
The lines correspond to the following models: FR (solid), DGP (dashed) and
$\Lambda$CDM (long dashed). In the insert panel we present 
the corresponding fractional difference 
of the DGP (dashed line) and $\Lambda$CDM (long dashed line) 
models with respect to the FR model. To produce the curves 
we use $\Omega_{m0}=0.27$.}
\end{figure}

In Fig.1 we present the evolution of the growth rate of clustering 
for the current cosmological models, ie., the
FR model (solid line), the DPG (dashed line) and standard 
$\Lambda$CDM (long dashed line) models
and the fractional difference between the first
(FR model) and each of the other two models (insert panel).
Notice that in order to produce the curves we utilize $\Omega_{m0}=0.27$.
The general behavior 
of the functional form of the FR growth rate 
is an intermediate case between the DGP and $\Lambda$CDM growth rates.

\begin{figure}[ht]
\mbox{\epsfxsize=8.5cm \epsffile{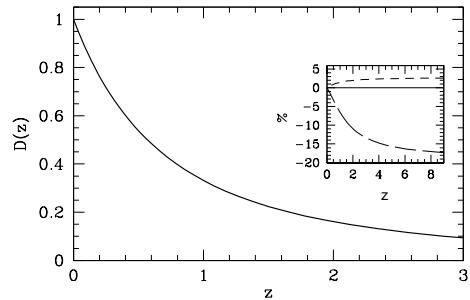}} 
\caption{
The evolution of the growth factor, with that
corresponding to
the FR model ($\gamma_{FR}=9/14$) showing a $\sim 0.1-2\%$
difference with respect to that of 
the DGP model ($\gamma_{DGP}=11/16$), especially at
large redshifts ($z\ge 1$). Notice that the growth factor 
normalized to unity at the present time 
}
\end{figure}

In Fig.2  we show the growth factor 
evolution which is derived by integrating 
Eq. (\ref{fzz221}), for the FR cosmological model.
In the insert panel of Fig.2 we plot the
fractional difference between the different models, similarly to
Fig.1, but now for the growth factor.
Obviously, the growth factor of the flat FR model 
is slightly different [$\frac{\delta D}{D_{FR}}(\%)\sim 0.1-2\%$]
from the one provided by the conventional flat DGP cosmology. 
Concering the $\Lambda$CDM model,
the expected differences are small at low redshifts, but become
gradually larger for $z\ge 1$, reaching variations of up to
$\sim -16\%$ at $z\sim 4$.

We would like to end this section with a brief discussion 
about the observational consequences of the FR model.
Since the flat FR model shares 
exactly the same Hubble parameter with 
the flat DGP model, this implies that the flat FR model 
inherits all the merits and demerits 
of the flat DGP gravity model. Thus, it becomes obvious that 
the FR model is under observational pressure when we compare against 
the background cosmological data (SNIa, BAO, CMB shift parameter). 
We would like to mention that the FR model is in agreement with 
the SNIa data \cite{Stav07}
(a similar situation holds also for the DGP, see Sec. II). 
As far as the ISW effect is concerned the situation is almost the same. 
In particular, the dependence of the ISW effect on the 
different cosmologies enters through the
different behavior of $D(a)$ (growth factor), which is affected 
by $\gamma$, and of $H(a)$ (see Eq. 14.16 in Ref. \cite{Ame10b}).
Taking the above arguments into account-namely, the same $H(a)$ and very 
small difference in $D(a)$ [see insert panel of Fig. 2], we conclude 
that both flat FR and DGP 
models predict almost the same ISW effect which of course is in disagreement 
with the ISW observational data. 
It is however possible to derive an extended version of the FR model 
free from the observational problems by including additional terms 
of the Finslerian metric $f_{\mu \nu}$ in the modified 
Friedmann equation. 
Such an analysis is in progress and will be published elsewhere.

\section{Conclusions}
In this paper we compared the Finsler-Randers (FR) space-time 
against the DGP gravity model. To our surprise, we found 
that the flat FR space-time is perfectly equivalent to the cosmic
expansion history of the flat DGP cosmological model, despite the fact that 
the two models live in a completely different geometrical background.
At the perturbative level we studied the 
linear growth of matter perturbations 
and it was found that the FR growth index is 
$\gamma_{FR} \simeq 9/14$ which is almost $\sim 7\%$ less than 
the theoretically predicted value of the DGP gravity model 
$\gamma_{DGP}\simeq 11/16$. The latter implies that 
the growth factor of the flat FR model is slightly different ($\sim 0.1-2\%$)
from the one provided by the conventional flat DGP cosmology.

\vspace {0.4cm}

{\bf Acknowledgments:}  
{\it We would like to thank the anonymous referee for useful 
comments and suggestions.}

\end{document}